%% file: main.tex
\documentclass[conference]{IEEEtran}

\usepackage[
  pass,
]{geometry}

\usepackage[latin1]{inputenc}
\usepackage{amsmath,bbm,comment,subfigure}

\usepackage{graphicx}
\usepackage{acronym}
\usepackage{changes}
\usepackage{lipsum}
\definechangesauthor[name={Authors}, color=blue]{authors}
\setremarkmarkup{(#2)}

\usepackage{acronym}

\usepackage[acronym]{glossaries}
\usepackage{glossary-mcols}
\usepackage{url}
\usepackage{booktabs}
\usepackage{flushend}

\usepackage{amsmath}
\usepackage{algorithm}
\usepackage[noend]{algpseudocode}
\usepackage{siunitx} 

\usepackage{array}
\newcolumntype{L}[1]{>{\raggedright\let\newline\\\arraybackslash\hspace{0pt}}m{#1}}
\newcolumntype{C}[1]{>{\centering\let\newline\\\arraybackslash\hspace{0pt}}m{#1}}
\newcolumntype{R}[1]{>{\raggedleft\let\newline\\\arraybackslash\hspace{0pt}}m{#1}}

\makeatletter
\def\BState{\State\hskip-\ALG@thistlm}
\makeatother

\makeglossaries
\input{acronyms} 

\def\tallafigura{0.3}
\def\tallafiguraS{0.3}

\newglossarystyle{modsuper}{%
  \glossarystyle{super}%
}

\begin{document}

\title{Higher aggregation of gNodeBs in Cloud-RAN architectures via parallel computing}


\author{
\IEEEauthorblockN{Veronica Quintuna Rodriguez and Fabrice Guillemin} 
\IEEEauthorblockA{Orange Labs,   2 Avenue Pierre Marzin, 22300 Lannion, France \\       \{veronica.quintunarodriguez,fabrice.guillemin\}@orange.com}
}

\maketitle  
\begin{abstract}
In this paper, we address  the virtualization and the centralization of real-time network functions, notably in the framework of Cloud RAN (C-RAN). We thoroughly analyze the required fronthaul capacity for the deployment of the proposed C-RAN architecture. We are specifically interested in the performance of the software based channel coding function. We develop a dynamic multi-threading approach to achieve parallel computing on a multi-core platform. Measurements from an OAI-based testbed show important gains in terms of latency; this enables the increase of the distance between the radio elements and the virtualized RAN functions and thus a higher aggregation of gNodeBs in edge data centers, referred to as Central Offices (COs). 

\end{abstract}

\noindent {\bf Keywords:} NFV, Cloud-RAN, gNodeB, BBU, channel coding, OAI, multi-core, scheduling.

\section{Introduction}

The next generation of mobile networks promises not only broadband communications and very high data rates but customized and optimized network services for specific vertical  markets (e.g, Health, Automotive, Media and Entertainment)~\cite{verticals}. 5G mobile networks consider heterogeneous \gls{RAN} architectures for targeting different types of mobile access (WiFi, cellular femto, small, and macro cells) and for fulfilling service requirements especially in  terms  of  latency,  resilience,  coverage,  and bandwidth.

In the perspective of achieving specific end-to-end service performances, the virtualization of network functions is highly desirable to flexibly deploy network services in cloud infrastructures according to customer needs. For example, new RAN architectures aim at virtualizing and centralizing higher-\gls{RAN} functions in the network while keeping lower-RAN functions in distributed units (near to antennas). These two nodes, respectively referred to as \gls{CU} and \gls{DU} by the 3GPP, enable flexible and scalable functional splits, which can be adapted to the required network performance. In addition, the collocation of \gls{CU} with Mobile/Multi-access Edge Computing facilities opens the door to the realization of low latency services, thus meeting the strict requirements of URLLC (Ultra Reliable Low Latency Communications) identified by the 3GPP~\cite{3GPP_uRLLC}.

Centralizing RAN functions higher in the network however raises two main issues: low latency processing of radio signals (namely, base-band processing) and high capacity fiber-links in the fronthaul network. These two issues are addressed in the present work.

Given that  channel coding processing (i.e., a physical-layer function) is the most consuming in terms of computing resources and also the most sensitive with regard to performance, in particular the robustness of selected codes against interference, it seems essential to keep this function in the \gls{CU}. Via resource pooling, it is possible to achieve statistical multiplexing in the utilization of cores of a multi-core platform and thus to gain economies of scale while guaranteeing the deadline compliance in the execution of encoding/decoding functions. In addition, a global view of channel coding for several \glspl{gNB} enables better radio resource management via the adaptation of coding to predictable interference and also \gls{CoMP} technologies, i.e., interference reduction and better throughput. 

In order to reduce  fiber bandwidth requirements, we address in this work a bi-directional intra-PHY functional split which transmits both encoded and decoded data, in the \gls{DL} and \gls{UL} directions, respectively over Ethernet. In addition, we increase the fronthaul transmission time budget by improving the execution time of RAN functions. 

The basic principle to accelerate RAN functions, i.e., to reduce latency, consists of parallelizing the coding and decoding functions, either on the basis of \glspl{UE} or \glspl{CB}. The \gls{CB} is the smallest coding unit, which can be individually handled by the coding/decoding function of the RAN. The parallelization principles of channel coding are described in ~\cite{rodriguez2017towards,rodriguez2017performance,jsac}. In this paper, we go one step forward and focus on the implementation of the proposed thread-based models on a \gls{COTS} multi-core server by modifying the \gls{OAI} \gls{eNB} (an open source solution)~\cite{oaiWebSite}. We  report performance measurements from this platform by connecting the \gls{eNB} to a second server supporting an OAI-based core network and by observing the traffic generated by \glspl{UE} (commercial smart-phones), which are connected to the \gls{eNB} via an USRP card.

We furthermore provide the required fronthaul bandwidth supporting the proposed C-RAN architecture and evaluate the various intra-PHY functional splits currently envisaged by 3GPP~\cite{3GPP38_801} and studied by eCPRI~\cite{eCPRI} and IEEE~\cite{IEEEP1914.3}.

The organization of this paper is as follows: In Section~\ref{related}, we review the various functional splits considered in the literature and different solutions to reducing the time necessary to execute RAN functions. In Section~\ref{sec:fronthaul}, we evaluate the bandwidth requirements for various functional splits and formulate a recommendation for the best option in our understanding. In Section~\ref{testbed}, we describe the implementation of the multi-threading approach for the channel coding function in an OAI open source \gls{eNB}. Performance results are reported in Section~\ref{performance}. Concluding remarks are presented in Section~\ref{conclusion}.








\section{Related work}
\label{related}

Forthcoming 5G standards consider the coexistence of several functional splits of C-RAN architectures. For instance, the 3GPP  proposes eight options for splitting the E-UTRAN protocol at different levels. Each functional split meets the requirement of specific services. The most ambitious one (namely, the \textit{PHY-RF split} which corresponds to  option $8$ of 3GPP TR 38.801 Standard~\cite{3GPP38_801}) aims at a high level of centralization and coordination and enables efficient resource management of both radio (e.g., pooling of physical resources, \gls{CoMP} technologies) and cloud resources (e.g, statistical multiplexing of the computing capacity). However, this configuration (here referred to as \textit{\gls{FS}-I}) brings some deployment issues, notably tight latency and high-bandwidth on fronthaul links.

In fact, there is an open debate concerning the adoption of the most appropriate fronthaul transmission protocol over fiber. The problem relies not only on the constant bit rate performed by the currently used \gls{CPRI}~\cite{duan2016performance} protocol but on the high redundancy present in the transmitted \gls{I/Q} signals. Many efforts are currently being devoted to reducing optic/fiber resource consumption such as \gls{I/Q} compression~\cite{guo2013lte}, non-linear quantization, sampling rate reduction among others. Incoming \gls{CPRI} variants, notably those proposed by Ericsson et al.~\cite{eCPRI} perform \gls{CPRI} packetization via IP or Ethernet. A similar approach to \gls{RoE} is being defined by the IEEE Next Generation fronthaul Interface (1914) working group~\cite{IEEEP1914.3}. It specifies the encapsulation of digitized radio \gls{I/Q} payload for both control and user data. The xRAN Forum, which gathers industrials and network operators, is also producing an open specification for the fronthaul interface~\cite{xRAN}. It considers intra-PHY splitting as defined by 3GPP in TR 38.801~\cite{3GPP38_801}. A detailed fronthaul capacity analysis is addressed in Section~\ref{sec:fronthaul}.

While numerous fronthaul solutions are being standardized, less attention  is paid by the industry and academia to the runtime latency of virtualized \gls{RAN} functions. First studies concerning the computing performance are presented in~\cite{nikaein2015processing}. Authors compare the processing time of different virtualized environments, namely \gls{LXC}, Docker and \gls{KVM}; however, the number of concurrent threads/cores per \gls{eNB} is limited to $3$ since  parallel processing of intra-sub-frame is not performed, i.e., a single-core is dedicated to the whole processing of an LTE sub-frame. 

On the contrary, the multi-threading model presented in~\cite{rodriguez2017performance} performs data parallelism at a finer granularity, which enables an important latency reduction. The authors carry out an in-depth analysis of the workload and data structures handled during the base-band processing of the radio signals for both the \gls{DL}~\cite{rodriguez2017vnf} and \gls{UL}~\cite{rodriguez2017towards} directions. Two multi-threading solutions are then proposed for the channel coding function, which is the most resource consuming. First, the sub-frame data is decomposed in smaller data structures so-called \gls{TB}, which can be executed in parallel. A \gls{TB} corresponds to the data of a single \gls{UE} scheduled within one millisecond. It turns out that the runtime of the channel coding function (i.e., encoding in the \gls{DL} and decoding in the \gls{UL}) is directly proportional to the \gls{TBS}. A finer breaking of sub-frames is also presented; it considers the execution of \gls{CB} in parallel. A \gls{CB} is the smallest data unit, which can be individually treated by the channel coding function. The behavior of both parallelism by \glspl{UE} and parallelism by \gls{CB} is evaluated by simulation and presents a gain up to $60\%$.  In Section~\ref{testbed}, we describe an implementation of the proposed schemes and give the performance results in Section~\ref{performance}.

\section{Fronthaul capacity}
\label{sec:fronthaul}

\subsection{Problem formulation}

One of the main issues of Cloud-RAN (C-RAN) is the required fiber bandwidth to transmit base band signals between the BBU-pool (namely, \gls{CU}) and each antenna (namely, \gls{DU} or \gls{RRH}). The fronthaul capacity is determined by the number of base band units (one per \gls{gNB}) hosted in the data center at the edge of the network (referred to as \gls{CO}). The current widely used protocol for data transmission between antennas and \glspl{BBU} is \gls{CPRI} which transmits \gls{I/Q} signals. The transmission rate is constant since \gls{CPRI} is a serial \gls{CBR} interface. It is then independent of the mobile network load~\cite{duan2016performance}. Several functional splits of the physical layer can then be analyzed in order to save fiber bandwidth~\cite{wubben2014benefits,duan2016performance}. The required fronthaul capacity for the various functional splits is presented  below.

\subsection{Required capacity per functional split}

\begin{figure*}[hbtp]
  \centering
  \includegraphics[scale=.53, trim=0 210 0 0, clip]{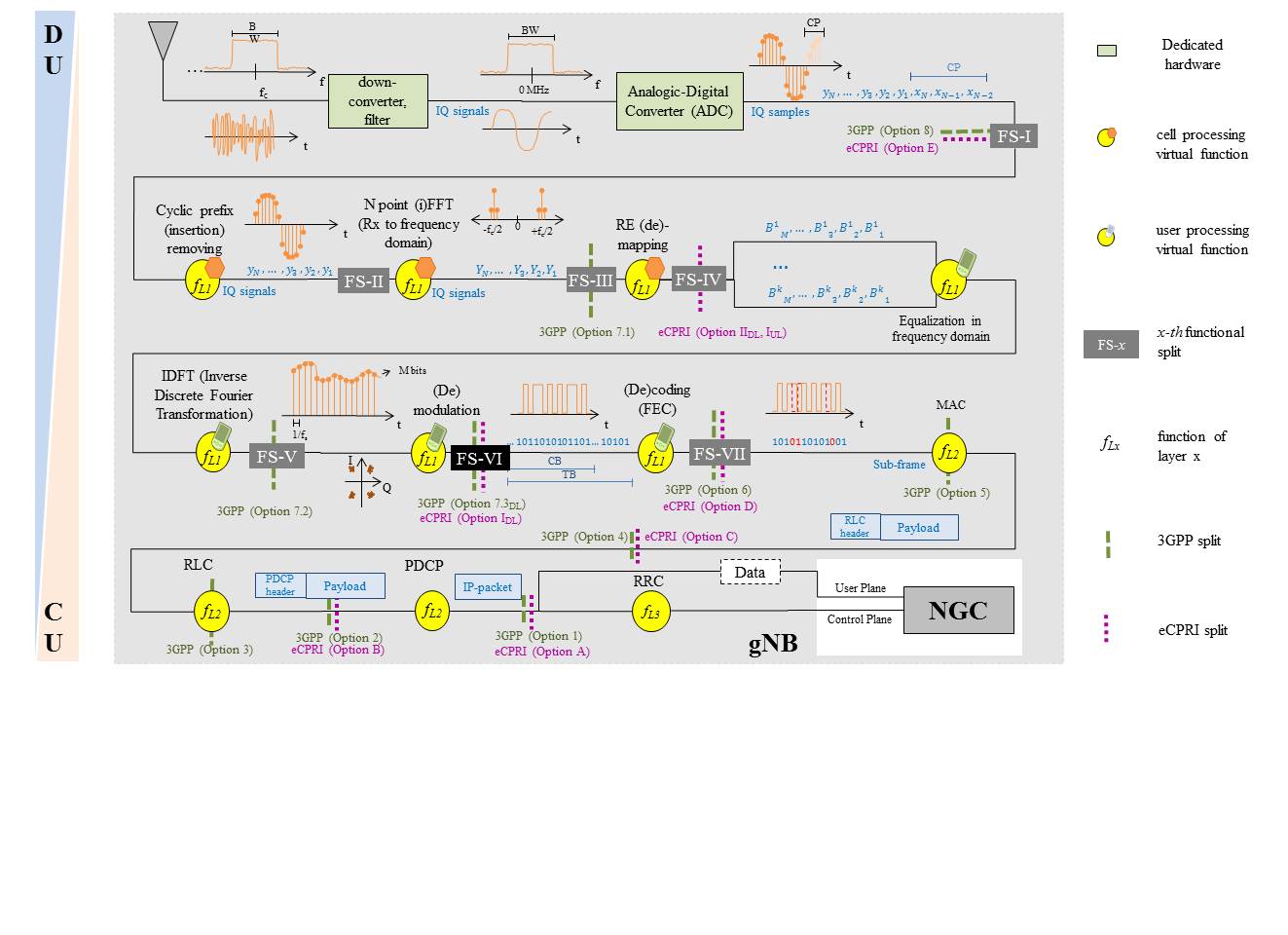}
    \caption{Fronthaul radio interfaces according to the various functional splits.}
  \label{fig:functionalsplits}
\end{figure*}

We have illustrated in Figure~\ref{fig:functionalsplits} the various functions executed in a classical RAN. For the downlink direction, IP data packets are first segmented by the PDCP and RLC layers. Then, the MAC layer determines the structure of the subframes (of 1~ms in LTE) forming  frames of 10~ms to be transmitted to \glspl{UE}. Once the MAC layer has fixed the allocation of \gls{PRB} for the \glspl{UE}, information is coded in the form of \glspl{CB}. Then, remaining L1 functions are executed on the encoded data for their transmission (modulation, Fourier transform, giving rise to \gls{I/Q} signals).  In the uplink direction, the functions are executed in reverse order.

The functional split actually defines the centralization level of RAN functions in the cloud-platform, i.e., it determines which functions are processed in \textit{dedicated} hardware near to antennas (\gls{DU}) and those which are moved higher in the network to be executed in centralized data centers (\gls{CU}). 

The required fronthaul capacity significantly decreases when the functional split is shifted after the PHY layer or even after the MAC layer~\cite{wubben2014benefits}.  It is worth noting that new RAN implementations consider the coexistence of configurable functional splits where each of them is tailored to the  requirements of a specific service or to a network slice. For instance, \gls{URLLC} expects a one-millisecond round-trip latency between the \gls{UE} and the \gls{gNB} while \gls{eMBB} requires only $4$ milliseconds. 

In the following, we shall pay special attention to C-RAN supporting fully centralization.
The required fronthaul capacity (given in Mbps) for all intra-PHY functional splits, (denoted, for short, by $FS-N$, $N$ ranging from 1 to 8) is presented in Table~\ref{tab:capacitySplits} as a function of the cell bandwidth $BW_{cell}$ (given in MHz).

\begin{table}[htbp]
\caption{Required fronthaul capacity in a Cloud-RAN system for various cell bandwidths.}
\label{tab:capacitySplits}
\begin{center}
\begin{tabular}{lrrrrrr}
\toprule
\textbf{$BW_{cell}$} & \textbf{$1.4$} & \textbf{$3$} &  \textbf{$5$} & \textbf{$10$} & \textbf{$15$} & \textbf{$20$} \\ 
\midrule
\gls{FS}-I  & $153.6$ & $307.2$ & $614.4$ & $1228.8$ & $1843.2$ & $2457.6$ \\ 
\gls{FS}-II  & $143.4$ & $286.7$ & $573.4$ & $1146.9$ & $1720.3$ & $2293.8$ \\ 
\gls{FS}-III & $86.4$ & $172.8$ & $360.0$ & $720.0$ & $1080.0$ & $1140.0$ \\ 
\gls{FS}-IV  & $60.5$ & $121.0$ & $252.0$ & $504.0$ & $756.0$ & $1008.0$ \\ 
\gls{FS}-V & $30.2$ & $60.5$ & $126.0$ & $252.0$ & $378.0$ & $504.0$ \\ 
\gls{FS}-VI & $6.0$ & $12.1$ & $25.2$ & $50.4$ & $75.6$ & $100.8$ \\ 
\gls{FS}-VII & $5.5$ & $11.1$ & $23.1$ & $46.2$ & $69.3$ & $92.4$ \\ 

\bottomrule
\multicolumn{7}{c}{ }  \\

\multicolumn{7}{c}{\scriptsize $BW_{cell}$ [MHz], ${FS}-x$ [Mbps].} \\

\end{tabular}
\end{center}
\end{table}

\subsubsection*{Functional Split I}

The fully centralized architecture (Option $8$ according to 3GPP), referred to in this paper as \gls{FS}-I, only keeps in the \gls{DU} the down-converter, filters and the \gls{ADC}. \gls{I/Q} signals are transmitted from and to the \gls{CU} by using the \gls{CPRI} standard.  The problem of \gls{FS}-I is in the fact that the required fronthaul capacity does not depend on the traffic in a cell but of the cell bandwidth. 
The required data rate per radio element is given by
\begin{align*}
 R_{1} &=2*M*f_s*F_{coding}*F_{control}*N_{ant} \\
       &=2*M*N_{FFT}*BW_{sc}*F_{coding}*F_{control}*N_{ant},
\end{align*}
where the various variables are defined in Table~\ref{tab:parameters}.

\begin{scriptsize}
\begin{table}[htbp]
\caption{List of parameters impacting the required fronthaul capacity.}
\label{tab:parameters}
\begin{center}
\begin{tabular}{L{1.5cm}L{3.5cm}L{2.5cm}}
\toprule
\textbf{Parameter} & \textbf{Description} & \textbf{Value}  \\ 
\midrule
$BW_{sc}$ & \scriptsize sub-carrier bandwidth & \scriptsize  $15$ KHz\\ 
$BW_{LTE}$ & \scriptsize  LTE bandwidth & \scriptsize $1.4,3,5,10,15,20$ MHz \\ 
$BW_{uf}$ & \scriptsize useful bandwidth & \scriptsize $18$ MHz ($20$MHz)  \\ 
$f_c$ & \scriptsize nominal chip rate & \scriptsize $3.84$ MHz\\ 
$f_s$& \scriptsize sampling frequency & \scriptsize e.g., $30.72$ MHz ($20$MHz)\\ 
$F_{coding}$ & \scriptsize coding factor & \scriptsize $10/8$ or $66/64$\\ 
$F_{control}$ & \scriptsize control factor & \scriptsize $16/15$ (CPRI)\\ 
$F_{oversampling}$ & \scriptsize oversampling factor & \scriptsize $~1.7$ \\ 
$k$ & \scriptsize code rate & \scriptsize e.g., $11/12$~\cite{lopez2011optimization} \\ 
$M$ & \scriptsize number of bits per sample & \scriptsize $15$\\ 
$N_{ant}$ & \scriptsize number of antennas for MIMO & \scriptsize e.g., $2x2$\\ 
$N_{FFT}$ & \scriptsize number of FFT samples per OFDM symbol & \scriptsize e.g., $2048$ ($20$MHz)\\ 
$N_{RB}$ & \scriptsize total number of resource blocks per subframe &  \scriptsize e.g., $100$ ($20$MHz) \\ 
$N_{sc}$ & \scriptsize total number of sub-carriers per subframe & \scriptsize  e.g., $1200$ ($20$MHz)\\ 
$N_{sc-pRB}$ & \scriptsize number of sub-carriers per resource block &  \scriptsize $12$ \\ 
$N_{sy-psl}$ & \scriptsize number of symbols per time slot &  \scriptsize $7$ (normal CP) \\ 
$N_{sy-psf}$ & \scriptsize number of symbols per subframe &  \scriptsize $14$ (normal CP) \\ 
$O_m$ & \scriptsize modulation order & \scriptsize  $2$-QPSK,$4$-16QAM,$6$-64QAM,$8$-256QAM\\ 
$\rho$ & \scriptsize RBs utilization (mean cell-load) & \scriptsize $0.7$\\ 
$R_{x}$ & \scriptsize data rate when using the x-th functional split & \\ 
$T_{CP}$ & \scriptsize average duration of a cyclic prefix & \scriptsize $4.76\mu$s (normal CP) \\ 
$T_{s}$ & \scriptsize symbol duration & \scriptsize $66.67\mu$s (normal CP)\\ 
$T_{UD-psl}$ & \scriptsize useful data duration per time slot & \scriptsize $466.67\mu$s (normal CP) \\ 
\bottomrule
\end{tabular}
\end{center}
\end{table}
\end{scriptsize}


In Table~\ref{tab:capacitySplits}, we have computed the fronthaul capacity $R_1$ in function of the cell bandwidth. As the cell bandwidth increases, the required front haul capacity per sector  can reach $2.4$~GBit/s. Since each site is generally equipped with three sectors, we can observe that the required bandwidth reaches prohibitive values for this functional split.

\subsubsection*{Functional Split II}
When implementing the \gls{CP} removal in the \gls{DU}, the fronthaul capacity can be reduced. This solution may experiment correlation problems due to the \gls{ISI} apparition. The required data rate for this functional split is given by
\begin{multline*}
    R_{2} = 2*M*N_{FFT}*{(T_s+T_{CP})^{-1}}*F_{coding} \\ *F_{control}*N_{ant},
\end{multline*}
where $T_{CP}$ is the average duration of a \gls{CP} in a radio symbol. $T_{CP}=(500[us]-T_s[us]*7)/7=4.76191$ microseconds.
$T_s=\frac{1}{BW_{sc}}=\frac{1}{15KHz}$. The useful data duration in a radio slot ($500$ microseconds) is given by $T_{UD-psl}=T_s*N_{sy-psl}=66.67*7=466.67$ microseconds. The resulting fronthaul capacity is given in Table~\ref{tab:capacitySplits}. The reduction in bandwidth requirement is rather small when compared with FS-I.

\subsubsection*{Functional Split III}

By keeping the \gls{FFT} function near to antennas the required fronthaul capacity can be considerably reduced. In this case,  radio signals are transmitted in the frequency domain from radio elements to the CU for the uplink and vice versa for the downlink. This solution prevents from the overhead introduced when sampling the time domain signal. The oversampling factor is given by $F_{oversampling}=\frac{N_{FFT}}{N_{sc}}=~1.7$, e.g., $F_{oversampling}=512/300=1.71$ for an LTE bandwidth of $5$MHz. The corresponding fronthaul bit rate is then given by
\begin{align*}
    R_{3} &= 2*M*N_{sc}*BW_{sc}*F_{coding}*F_{control}*N_{ant} \\
          &= 2*M*N_{sc}*(T_s)^{-1}*F_{coding}*F_{control}*N_{ant} 
\end{align*}
As illustrated in Table~\ref{tab:capacitySplits}, the fronthaul capacity is halved when compared with the initial CPRI solution. In the following, we show that the fronthaul capacity can still be reduced by a factor $10$.

\subsubsection*{Functional Split IV}
When including the de-mapping process in the \gls{DU}, it is possible to adapt the bandwidth as a  function of the traffic load in the cell, then the required fronthaul capacity is directly given by the fraction of utilized radio resources~\cite{wubben2014benefits}. 

Here, only the \glspl{RB}, which carry information are transmitted. When considering the behavior of current deployed eNBs of the Orange mobile network serving a high-density zone (e.g, a train station), an eNB presents in average a \gls{RB} utilization of $11.96\%$ and $20.69\%$ in the uplink and downlink directions, respectively. The highest utilization values observed in current deployed Orange's \glspl{eNB} do not exceed $70\%$ in the downlink direction. Thus, if we take $\rho=0.7$ as the mean cell-load (worst-case), this yields a fronthaul bit rate equal to
\begin{equation*}
    R_{4} = 2*M*N_{sc}*BW_{sc}*F_{coding}*F_{control}*N_{ant}*\rho
\end{equation*}
Numerical values given in Table~\ref{tab:capacitySplits} show that the gain with respect to the previous solution is however rather small.

\subsubsection*{Functional Split V}

This configuration presents a gain in the fronthaul load, when \gls{MIMO} schemes are performed. The equalization function combines signals coming  from multiple antennas; as a consequence, the required fronthaul capacity is divided by $N_{ant}$. The required front haul capacity is then given by 
\begin{equation*}
    R_5 = 2*M*N_{sc}*BW_{sc}*F_{coding}*F_{control}*\rho
\end{equation*}
Table~\ref{tab:capacitySplits} shows a drop by a factor $2$ with respect to the previous solution.

\subsubsection*{Functional Split VI}
By keeping the demodulation/modulation function near to antennas, the required data rate is given by
\begin{equation*}
    R_{6} = N_{sc}*N_{sy-psf}*O_m,
\end{equation*}
where $N_{sy-psf}$ is the number of symbols per subframe (i.e., $N_{sy-psf}=14$ when using normal cyclic prefix),  $O_m$ is the modulation order, i.e., the number of bits per symbol. Taking the highest modulation order currently supported in the deployed networks, i.e, $O_m=6$, the required fronthaul capacity is reduced  to $100$ Mbps. This represents a significant gain when compared to the initial CPRI (FS-I) solution. It is also worth noting that this solution preserves the gain achievable by C-RAN.

\subsubsection*{Functional Split VII}

Just for the sake of completeness, we consider now the case when keeping the channel coding function near to antennas, redundancy bits are not transmitted. Nevertheless this configuration reduces the advantages of C-RAN. \glspl{DU} become more complex and expensive. The required fronthaul capacity is 
\begin{equation*}
    R_{7} = N_{sc}*N_{sy-psf}*O_m*k,
\end{equation*}
where $k$ is the code rate, i.e., the ratio between the useful information and the transmitted information including redundancy.  In LTE code rate $k$ commonly ranges from $1/12$ to $11/12$~\cite{lopez2011optimization}. In Table~\ref{tab:parameters}, we use $k=11/12$ as the worst-case.

\subsection{Functional split selection}

In view of the analysis carried out in the previous section, functional split VI seems to be the most appropriate.  It is then necessary to encapsulate the fronthaul payload within Ethernet, i.e., distributed units are connected to the centralized ones through an Ethernet network. RoE is considered by IEEE Next Generation fronthaul Interface (1914) Working Group as well as by the xRAN fronthaul Working Group of the xRAN Forum.

The main issue of an Ethernet-based fronthaul is the latency fluctuation~\cite{chih2015rethink}. Transport jitter can be isolated by a buffer, however, the maximum transmission time is constrained by the processing time of centralized functions. The sum of both transmission and processing time must meet \gls{RAN} requirements (i.e., $1$~ms for DL and $2$~ms for UL).

The transmission time can quickly rise due to the distance and the added latency at each hop (e.g., switches) in the network. The transmission time can be roughly obtained from the light-speed in the optic-fiber (e.g., $2.1$x$10^8$ m/s), and latency of $50\mu$s by hop~\cite{chih2015rethink}. For instance, the required transmission time for an \gls{gNB} located $40$ km from the \gls{CO} rises $280\mu$s$+50*8=680\mu$s. Hence, the remaining time-budget for BBU processing is barely $320\mu$s in the down-link direction. The proof-of-concept of C-RAN acceleration for supporting FS-VI is described below.

\section{Testbed and parallelization of coding functions}
\label{testbed}

\subsection{Testbed description}

\begin{figure}[hbtp] 
   \begin{center}
 \includegraphics[scale=\tallafigura, trim=0 360 0 0, clip] {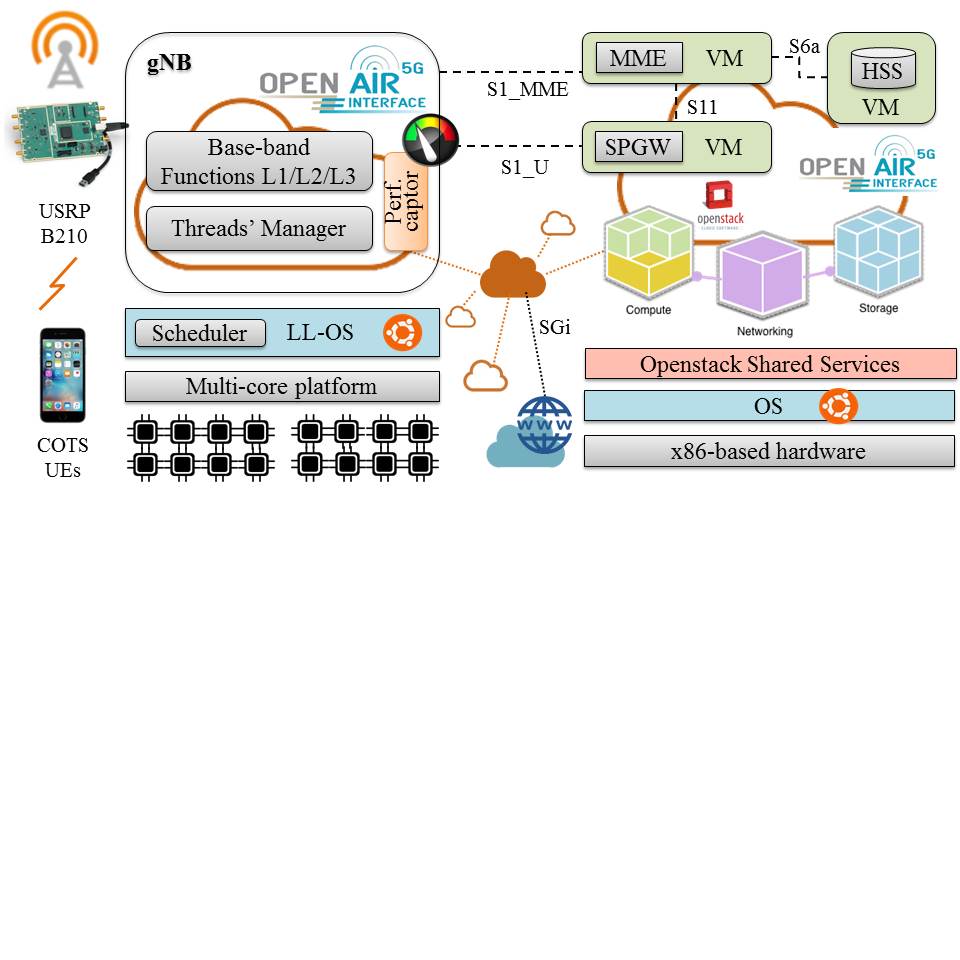}
   \caption{Testbed architecture.} 
   \label{fig:testbed}
\end{center}
\end{figure}

To evaluate the proposed C-RAN acceleration method and notably the gain in terms of latency when parallelizing  the channel coding, we have set up a testbed basically composed of $2$ servers (COTS PCs), one supporting the OAI EPC (MME, HSS, SPGW) and another  equipped with an USRP card and implementing a modified version of the OAI \gls{eNB} software.  This testbed is illustrated in Figure~\ref{fig:testbed}.

The platform implements a pool of threads, which perform the parallel processing of both encoding (downlink) and decoding (uplink) functions on a multi-core server. The workload of threads is managed by a global non-preemptive scheduler (so-called, thread manager); a thread is assigned to a dedicated single core with real-time OS priority and is executed until completion without interruption. The isolation of threads is provided by a specific configuration performed in the OS, which prevents from the use of channel coding computing resources for any other job.

\subsection{Implementation}

The goal of our modification of OAI code is to  perform massive parallelization of channel encoding and decoding processes. These functions are detailed  below, before presenting the multi-threading mechanism and the scheduling algorithm.

\subsubsection*{Encoding function}

\begin{figure}[hbtp]    
   \begin{center}
 \includegraphics[scale=\tallafigura, trim=0 255 0 0, clip] {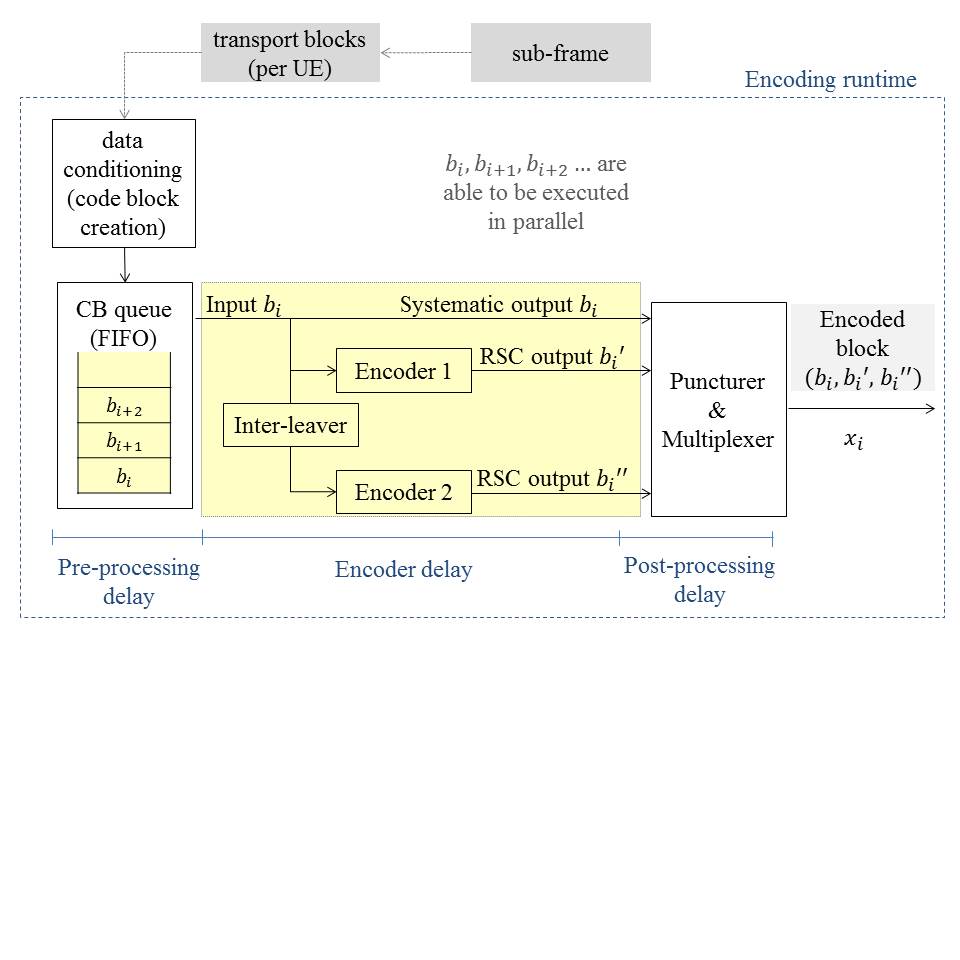}
   \caption{Block diagram of encoding function.} 
   \label{fig:encoding_block}
\end{center}
\end{figure}

The encoder (See Figure~\ref{fig:encoding_block} for an illustration) consists of 2 \gls{RSC} codes separated by an inter-leaver. Before encoding, data (i.e., a subframe) are conditioned and segmented in code blocks of size $T$, which can be encoded in parallel. When the multi-threading model is not implemented, \glspl{CB} are executed in series under a \gls{FIFO} discipline. Thus, an incoming data block $b_i$ is twice encoded, where the second encoder is preceded of the permutation procedure (inter-leaver). The encoded block $(b_i,b_i',b_i'')$ of size $3T$ constitutes the information to be transmitted in the downlink direction. Hence,  for each information bit two parity bits are added, i.e., the resulting code rate is given by $r=1/3$. With the aim of reducing the channel coding overhead, a puncturing procedure may be activated for periodically deleting bits. A multiplexer is finally employed to form the encoded block $x_i$ to be transmitted. The multiplexer is nothing but a parallel to serial converter which concatenates the systematic output $b_i$, and both recursive convolutional encoded output sequences, $b_i'$, and $b_i''$.

\subsubsection*{Decoding function}

Unlike encoding, the decoding function is iterative and works with soft bits (real and not binary values). Real values represent the \gls{LLR}, i.e., the radio of the probability that a particular bit was 1 and the probability that the same bit was 0 (log is used for better precision). 

The decoding function runs as follows: Received data $R(x_i)$ is firstly de-multiplexed in $R(b_i)$, $R(b_i)'$, and $R(b_i)''$, which  correspond to the systematic information bits of $i$-th code block $b_i$ and to the received parity bits $b_i'$ and $b_i''$, respectively.

\begin{figure}[hbtp]    
   \begin{center}
 \includegraphics[scale=\tallafigura, trim=0 240 0 0, clip] {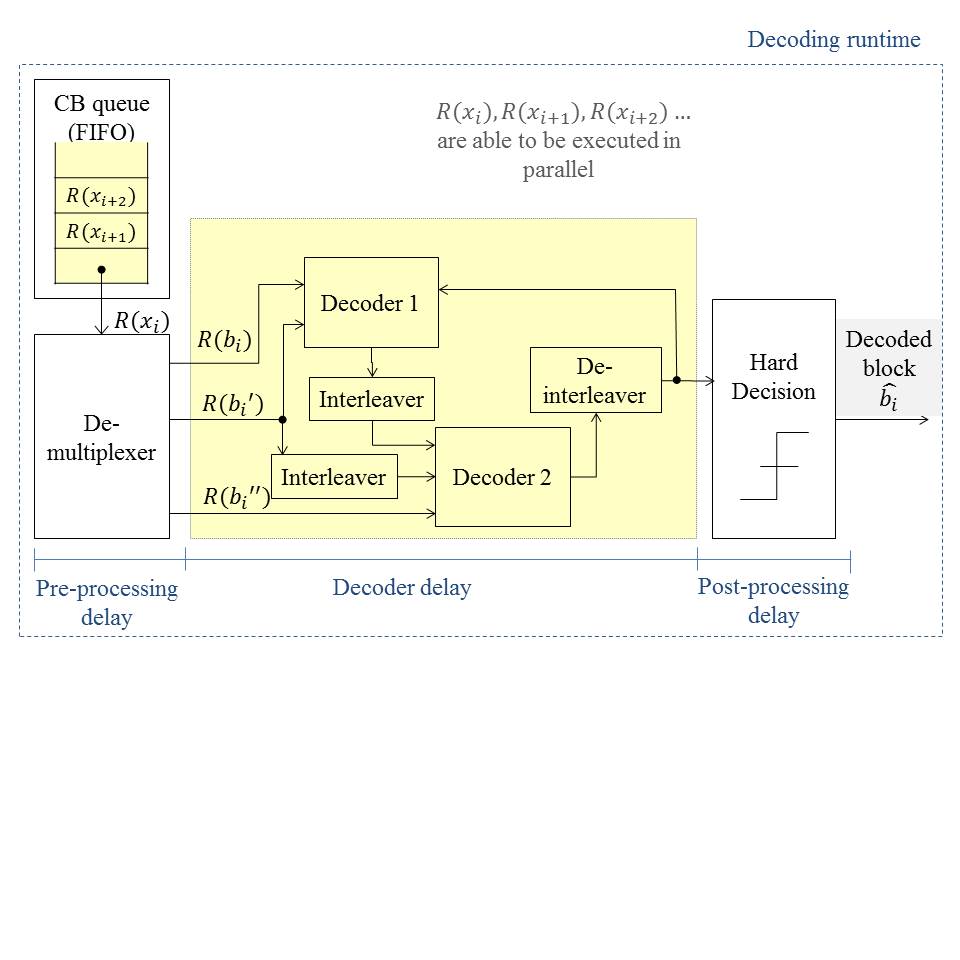}
   \caption{Block diagram of decoding function.} 
   \label{fig:decoding_block}
\end{center}
\end{figure}

$R(b_i)$ and $R(b_i)'$ feed the first  decoder  which calculates  the  \gls{LLR} (namely, extrinsic information)  and  passes  it  to the second decoder.  The second decoder uses that value to  calculate LLR and feeds back it to  the  first  decoder after a de-interleaved process. Hence, the second decoder has three inputs, the extrinsic information (reliability value) from the first decoder, the interleaved received systematic information $R(b_i)$, and the received values parity bits $R(b_i)''$. See Figure~\ref{fig:decoding_block} for an illustration.

The decoding procedure iterates until either the final solution is obtained or the allowed maximum number of iterations is reached. At termination, the final decision (i.e., $0$ or $1$ decision) is taken to obtain the decoded data block $\widehat{x_i}$. The data block  is either successfully  decoded  or   not.  The stopping criterion corresponds to the average mutual information of \gls{LLR}; if it converges the decoding process may terminate earlier. Note that there is a trade-off between the runtime (i.e., number of iterations) and the successful decoding of a data block.

\subsubsection*{Thread-pool}

On the basis of massive parallel programming, we propose splitting the channel encoding and decoding function in multiple parallel runnable jobs. The main goal is to improve their performance in terms of latency. 

\begin{figure}[hbtp]
     
   \begin{center}
 \includegraphics[scale=\tallafigura, trim=0 185 0 0, clip] {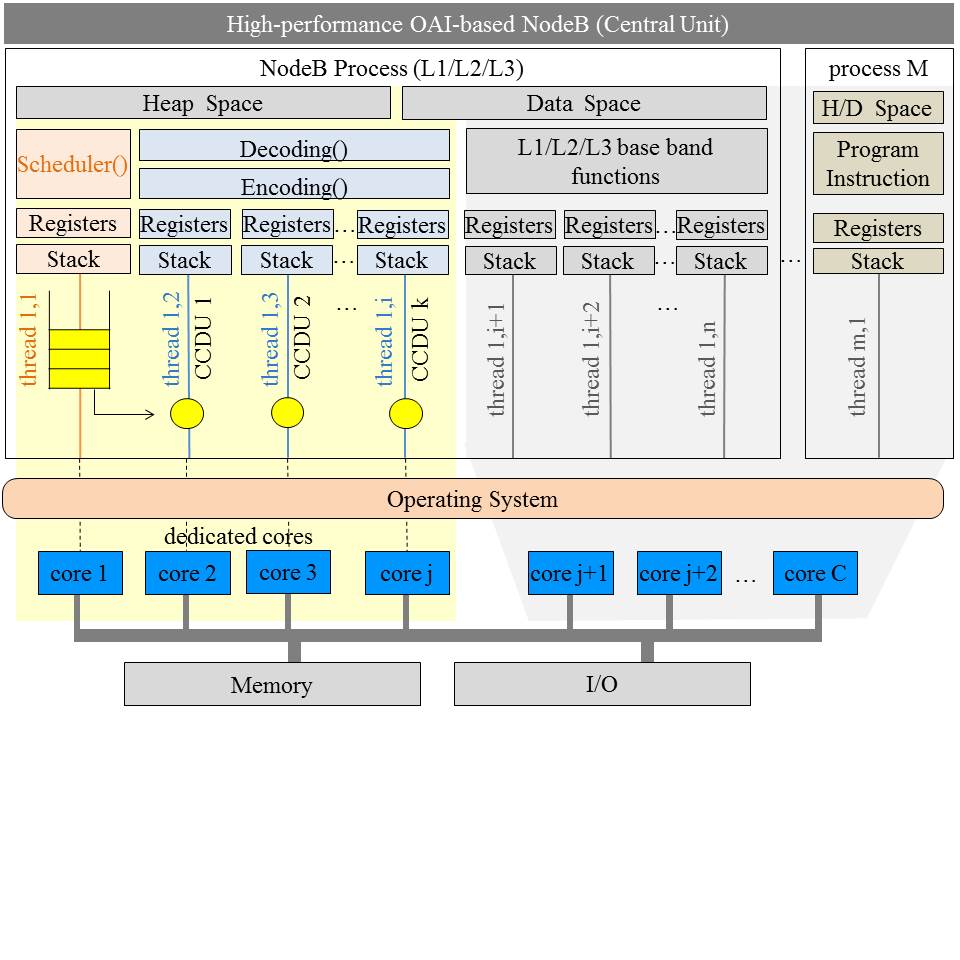}
   \caption{Multi-threading implementation.} 
   \label{fig:arch}
\end{center}
\end{figure}

In order to deal with the various parallel runnable jobs, we implement a thread-pool, i.e., a multi-threading environment. A dedicated core is affected to each thread during the channel coding processing.  When the number of runnable jobs exceeds the number of free threads, jobs are queued. 

To achieve low latency, we implement multi-threading within a single process instead of multitasking across different processes (namely, multi-programming). In a real-time system, creating a new process on the fly becomes extremely expensive because all data structures must be allocated and initialized. In addition, in a multi-programming \glspl{IPC} go through the \gls{OS}, which produces system calls and context switching overhead.

When using a multi-threading (namely, POSIX~\cite{butenhof1997programming}) process for running encoding and decoding functions, other processes cannot access resources (namely, data space, heap space, program instructions), which are reserved for channel coding processing.

The memory space is shared among all threads belonging to the channel coding process, which enables latency reduction. Each thread performs the whole encoding or decoding flow of a single \gls{CCDU}. We define a \gls{CCDU} as the suite of bits, which corresponds to a radio sub-frame (no-parallelism), a \gls{TB} or even a \gls{CB}. When performing parallelism, \glspl{CCDU} arrive in batches every millisecond. These data units are appended to a single queue (see Algorithm~\ref{algo:queuing}), which is managed by a global scheduler. We use non-preemptive scheduling, i.e., a thread (\gls{CCDU}) is assigned to a dedicated single core with real-time \gls{OS} priority and is executed until completion without interruption.

Isolation of threads is not provided by the POSIX API; hence, a specific configuration has been set up in the \gls{OS} to prevent the use of channel coding computing resources for any other jobs. The global scheduler (i.e., the thread manager) runs itself within a dedicated thread and performs a FIFO discipline for allocating cores to \gls{CC} jobs, which are waiting in the queue to be processed. Figure~\ref{fig:arch} illustrates $j$ cores dedicated to channel coding processing, remaining $C-j$ cores are shared among all processes running in the system, including those belonging to the upper-layers of the \gls{gNB}.

{

\begin{algorithm}
\caption{Queuing channel coding jobs}\label{algo:queuing}
\begin{algorithmic}[1]
\begin{scriptsize}
\State $CB\_MAX\_SIZE \gets 6120$
\Procedure{Queuing}{}
\While {$subframe\_buffer \neq \emptyset$} 
	\State $SF \gets \textit{get subframe}$
  \State $nUE \gets \textit{get UE's number}$    
    \While {$\mbox{nUE} > \mbox{0}$}
    	\State $CCDU_{UE} \gets \textit{get\_TB(nUE-th,SF)}$
        \If {$CB\_parallelism\_flag=true$}
            \While {$CCDU\_UE \geq CB\_MAX\_SIZE$}
        		\State $CCDU\_CB \gets \textit{get\_CB(nCB-th,CCDU\_UE)}$
                \State $queue \gets \textit{append(CCDU\_CB)}$
            \EndWhile
        \Else
        	\State $queue \gets \textit{append(CCDU\_UE)}$
        \EndIf  
		
        \State $nUE \gets \text{nUE-1}$
    \EndWhile
\EndWhile
\EndProcedure
\end{scriptsize}
\end{algorithmic}
\end{algorithm}
}

\subsection{Queuing principles}

The \gls{CCDU}'s queue is a chained list containing the pointers to the first and last element, the current number of \glspl{CCDU} in the queue and the \textit{mutex} (namely, mutual exclusion) signals for managing shared memory. The \textit{mutex} mechanism is used to synchronize access to memory space in the case when more than one thread require writing at the same time. In order to reduce waiting times, we perform data context isolation per channel coding operation, i.e., dedicated \gls{CC} threads do not access any global variable of the \gls{gNB} (referred to as `soft-modem' in \gls{OAI}). 

\subsubsection*{Scheduler}

The scheduler takes from the queue the next \gls{CCDU} to be processed and updates the counter of jobs (i.e., decrements the counter of remaining \glspl{CCDU} to be processed). The next free core executes the first job in the queue.

In the case of decoding failure, the scheduler purges all \glspl{CCDU} belonging to the same \gls{UE} (\gls{TB}). In fact, a \gls{TB} can be successfully decoded only when all \glspl{CB} have been individually decoded (See Algorithm~\ref{algo:scheduler}).

Channel coding variables are embedded in a permanent data structure to create an isolated  context per channel coding operation; in this way, \gls{CC} threads do not access any memory variable in the main soft-modem (\gls{gNB}). The data context is passed between threads by pointers.

\begin{algorithm}
\caption{Thread pool manager}\label{algo:scheduler}
\begin{algorithmic}[1]
\begin{scriptsize}
\Procedure{Scheduling}{}
\While {true}	
	\If {$queue=\emptyset$}
		\State \textit{wait next event}
    \Else
    	\State $CCDU\gets \textit{pick queue}$
    	\State \textit{process(CCDU)}
        	 \If {$decoding_failure=true$} 
            	\State \textit{purge waiting CCDU of the same TB}
             \EndIf 
         \State \textit{acknowledge(CCDU) done}
	\EndIf
\EndWhile
\EndProcedure
\end{scriptsize}
\end{algorithmic}
\end{algorithm}

\subsection{Performance captor}

In order to evaluate the performance of multi-threading, we have implemented a \emph{performance captor} which gets key timestamps during the channel coding processing for uplink and downlink directions.  With the aim of minimizing measurements overhead, data is collected by a separate process, so-called \emph{measurements collector}, which works out of the real-time domain. 

The data transfer between both separate processes, i.e., the \emph{performance captor} and the \emph{measurements collector}, is performed via an \gls{OS}-based pipe (also referred to as \textit{named pipe} or \textit{FIFO pipe} because the order of bytes coming in is the same as the order of bytes going out~\cite{bovet2005understanding}). 

Timestamps are got at several instants in order to obtain the following \glspl{KPI}: 
\begin{itemize}
\item Pre-processing delay, which includes data conditioning, i.e., code block creation, before triggering the channel coding itself. 
\item Channel coding delay, which measures the runtime of the encoder (decoder) process in the downlink (uplink).
\item Post-processing delay,  including the combination of \glspl{CB}. 
\end{itemize}

Collected traces contain various performance indicators such as the number of iterations carried out by the decoder per \gls{CB} as well as the identification of cores affected for both encoding and decoding processes. Decoding failures are detected when a value greater than the maximum number of allowed iterations is registered. As a consequence, the loss rate of channel coding processes as well as the individual workload of cores can be easily obtained.

\section{Performance results}
\label{performance}

In order to evaluate the performance of the proposed multi-threading scheme, we use the above described test-bed which contains a multi-core server hosting the \gls{gNB}. The various \glspl{UE} perform file transfers in both uplink and downlink directions. 

The test scenario is configured as follows: number of cells: $1$ \gls{gNB}; transmission mode: FDD; maximum number of RB: $100$; available physical cores: $16$; channel coding dedicated cores: $6$; number of \glspl{UE}: $3$.

 \begin{figure}[hbtp]
     
   \begin{center}
\subfigure[decoding (Rx)\label{fig:decoding_plat}]{{\includegraphics[scale=\tallafiguraS, trim=0 325 290 0, clip]{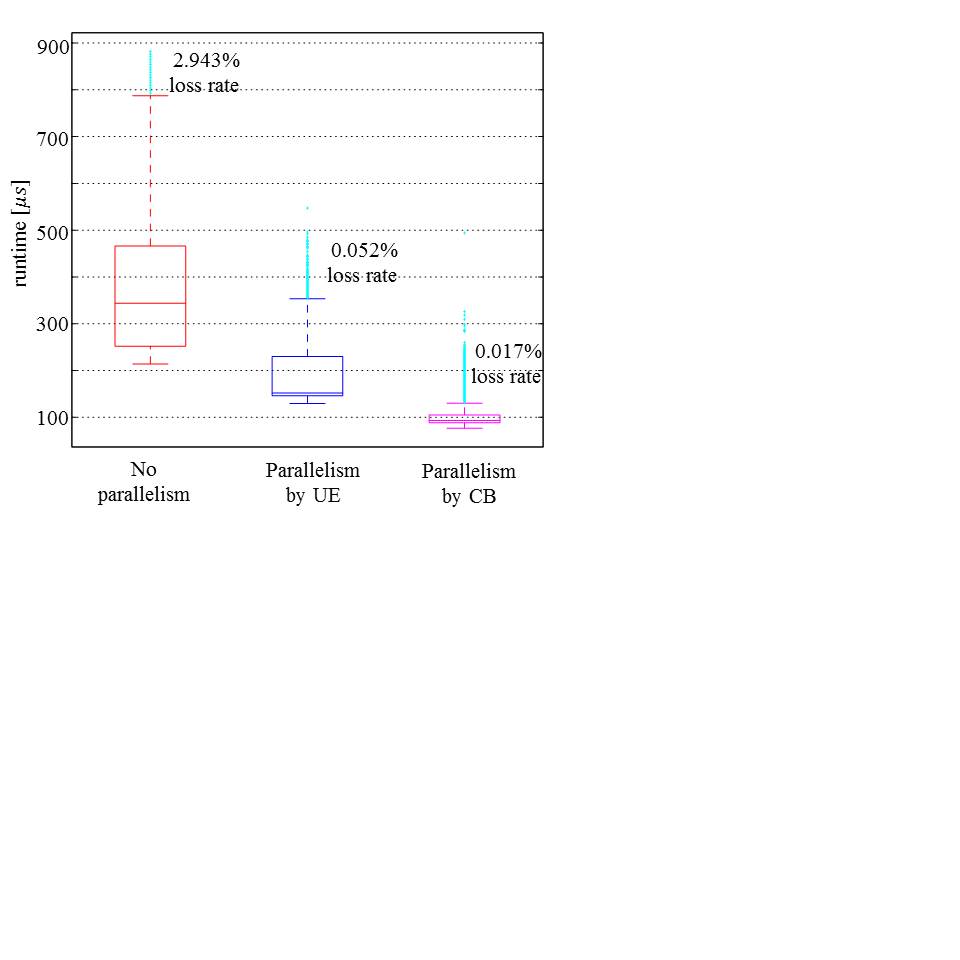}}}
\subfigure[decoder (Rx)\label{fig:decoder_plat}]{{\includegraphics[scale=\tallafiguraS, trim=0 325 290 0, clip]{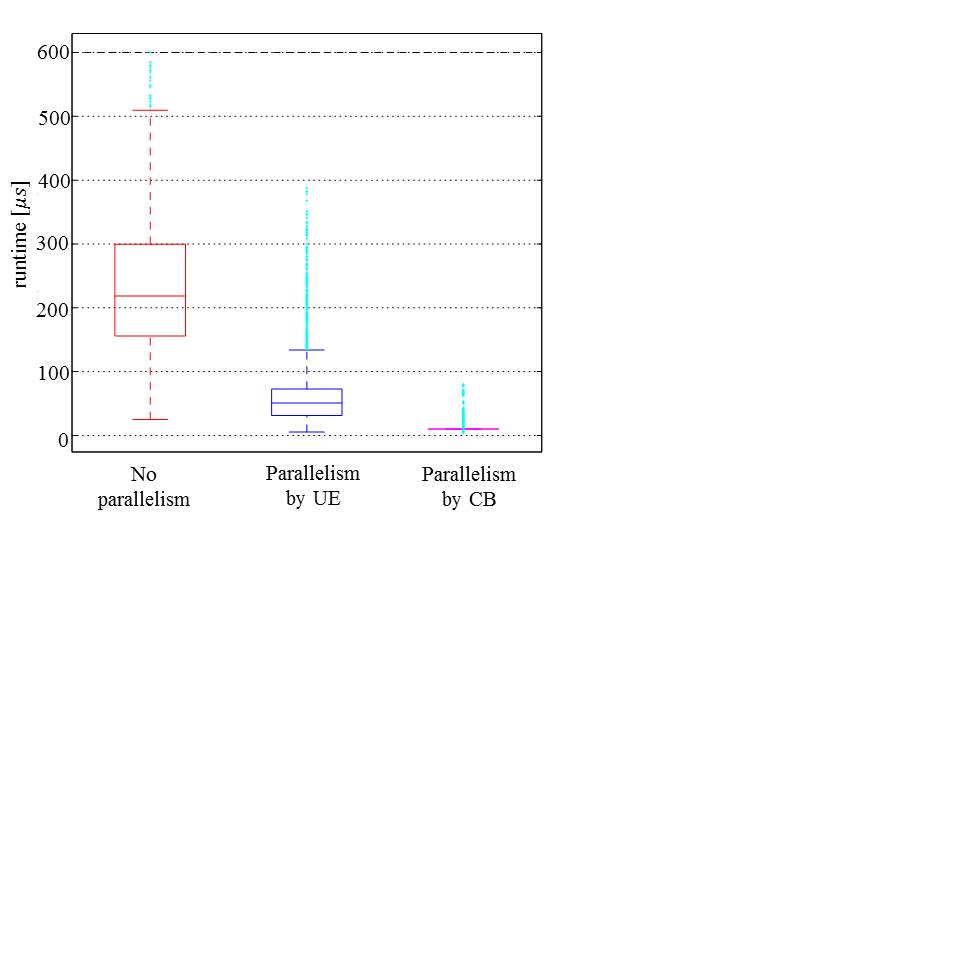}}}  
   \caption{Decoding runtime (test-bed).}    
   \label{fig:PoC_decoding}
\end{center}
\end{figure}

The performance captor takes multiple timestamps in order to evaluate the runtime of the encoder/decoder itself, as well as the whole execution time performed by the encoding/decoding function, which includes pre- and post-processing delays, e.g., code block creation, segmentation, assembling, decoder-bits conditioning (log-likelihood).  When a given data-unit is not able to be decoded, i.e., when the maximum number of iterations is achieved without success, data is lost and needs to be retransmitted. This issue is quantified by the KPI referred to as \textit{loss rate}. 

\begin{figure}[hbtp]
     
   \begin{center}
\subfigure[encoding (Rx)\label{fig:encoding_plat}]{{\includegraphics[scale=\tallafiguraS, trim=0 325 290 0, clip]{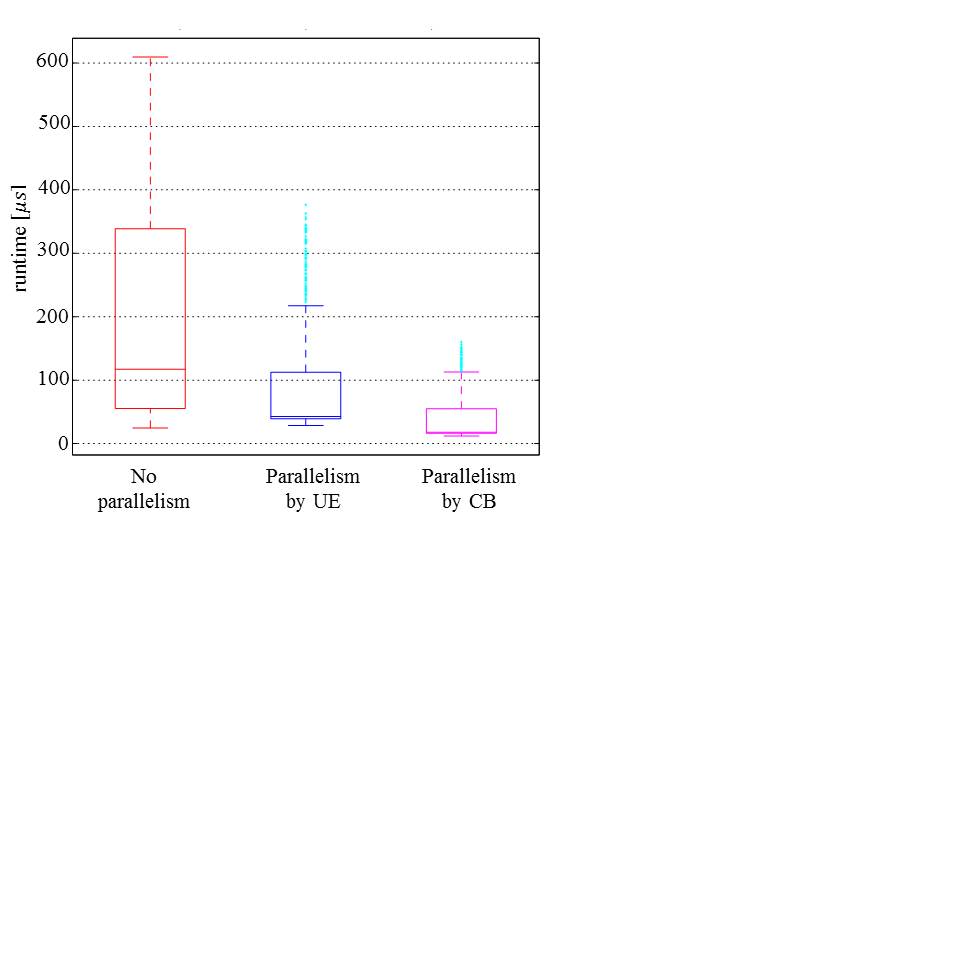}}}
\subfigure[encoder (Rx)\label{fig:encoder_plat}]{{\includegraphics[scale=\tallafiguraS, trim=0 325 290 0, clip]{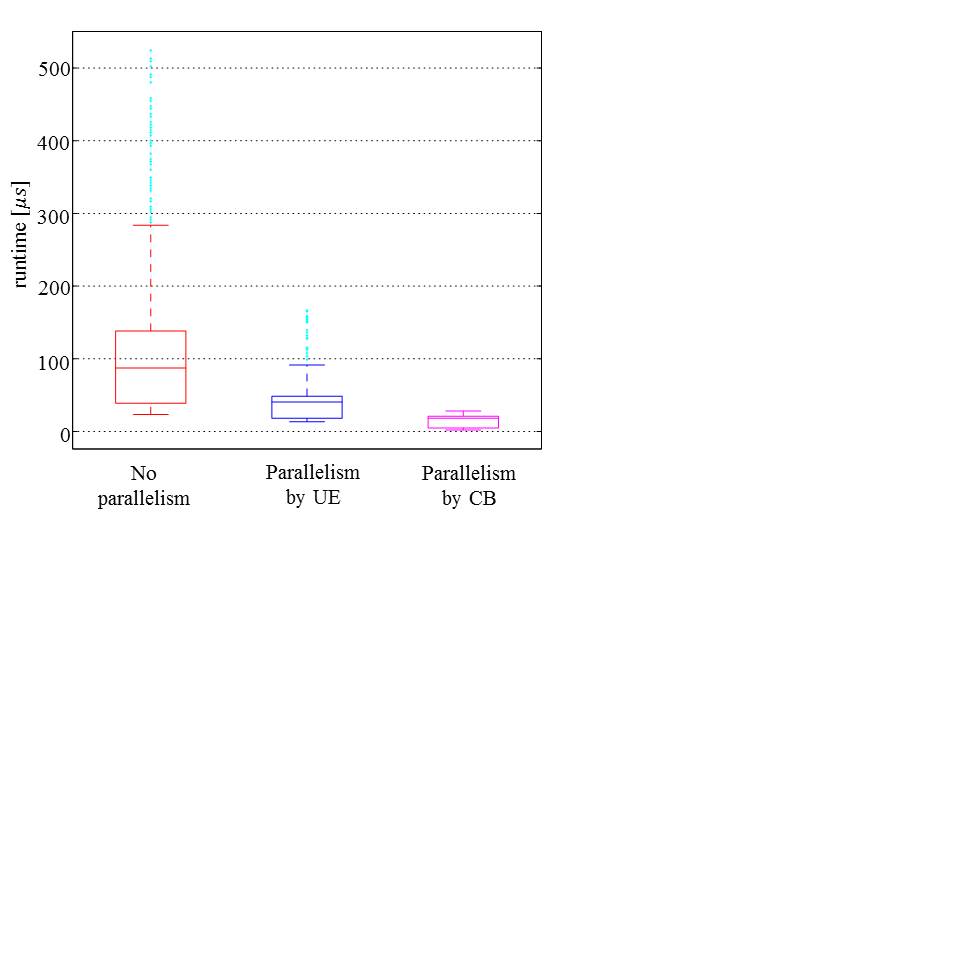}}}  
   \caption{Encoding runtime (test-bed).}    
   \label{fig:PoC_encoding}
\end{center}
\end{figure}

Runtime results are presented in Figures~\ref{fig:PoC_decoding} and~\ref{fig:PoC_encoding} for the uplink and downlink directions, respectively.  
Decoding function shows a performance gain of $72,6\%$ when executing \glspl{CB} in parallel, i.e., when scheduling jobs at the finest-granularity. Beyond the important latency reduction, runtime values present less dispersion when performing parallelism, i.e., runtime values are concentrated around the mean especially when executing \glspl{CB} in parallel. This fact is crucial when dimensioning cloud-computing infrastructures and notably data centers hosting virtual network functions with real-time requirements. When considering the gap between CB-parallelism and no-parallelism maximum runtime values, the C-RAN system (also referred to as BBU-pool) may be moved several tens of kilometers higher in the network. 

\section{Conclusion}
\label{conclusion}

In this work, we have addressed the two main issues of C-RAN architectures, i.e., high fronthaul capacity for transmitting radio signals between the radio elements and the Central Office, and tight latency for the execution of virtualized RAN functions in general purposes servers. 

In the aim of taking advantage of the benefits of C-RAN systems (i.e., spectral efficiency, interference reduction, data rate improvement) we focus on the study of \textit{fully centralized RAN architectures} which notably include the channel coding function in the CU. We have thus proposed an bi-directional intra-PHY functional split (namely, FS-VI) for transmitting encoded and decoded data over Ethernet. 

For meeting low latency requirements, we have performed C-RAN acceleration by means of parallel processing of the channel coding function. We concretely implemented on the basis of various open-source solutions, an end-to-end virtualized mobile network, which notably comprises a virtualized RAN. The platform particularly implements a thread-pool and two scheduling strategies, namely, parallelism by \glspl{UE} and parallelism by \glspl{CB}.  The parallel processing of both encoding (downlink) and decoding (uplink) functions is carried out in a multi-core server within a single OS process in order to avoid multi-tasking overhead. Results show important gains in terms of latency, which opens the door for deploying fully centralized \textit{cloud-native} RAN architectures.

\bibliographystyle{unsrt}
\bibliography{biblio}

\noindent

\end{document}

%% file: acronyms.tex
\newacronym{AAA}{AAA}{Authentication, Authorization, and Accounting}
\newacronym{RSC}{RSC}{Recursive  Systematic Convolutional}
\newacronym{LLR}{LLR}{Log-Likelihood Ratio}
\newacronym{FS}{FS}{Functional Split}
\newacronym{BBU}{BBU}{Base Band Unit}
\newacronym{COTS}{COTS}{Commercial off-the-shelf }
\newacronym{VNF}{VNF}{Virtualized Network Function}
\newacronym{VNF FG}{VNF FG}{VNF Forwarding Graph}
\newacronym{NFV}{NFV}{Network Function Virtualization}
\newacronym{GPP}{GPP}{General Purpose Processor}
\newacronym{vEPC}{vEPC}{virtual Evolved Packet Core}
\newacronym{LTE}{LTE}{Long Term Evolution}
\newacronym{URLLC}{URLLC}{Ultra-Reliable Low-Latency Communications}
\newacronym{eMBB}{eMBB}{enhanced Mobile Broad-Band}
\newacronym{MNO}{MNO}{Mobile Network Operator}
\newacronym{CDC}{CDC}{Centralized Data Center}
\newacronym{ATM}{ATM}{Asynchronous transfer mode}
\newacronym{HDR}{HDR}{High Data Rate}
\newacronym{NBS}{NBS}{Nash Bargaining Solution}
\newacronym{C-EPC}{C-EPC}{Cloud-EPC}
\newacronym{EPCaaS}{EPCaaS}{EPC as a Service}
\newacronym{TDD}{TDD}{Time Division Duplex}
\newacronym{UE}{UE}{User Equipment}
\newacronym{HARQ}{HARQ}{Hybrid Automatic Repeat-Request}
\newacronym{PRB}{PRB}{Physical Resource Blocks}
\newacronym{MCS}{MCS}{Modulation and Coding Scheme}
\newacronym{CQI}{CQI}{Channel Quality Indicator}
\newacronym{DC}{DC}{Dedicated Core}
\newacronym{RR}{RR}{Round Robin}
\newacronym{G}{G}{Greedy}
\newacronym{SNR}{SNR}{Signal Noise Ratio}

\newacronym{FDD}{FDD}{Frequency Division Duplex}
\newacronym{OFDM}{OFDM}{Orthogonal Frequency Division Multiplexing}

\newacronym{VM}{VM}{Virtual Machine}
\newacronym{PDCP}{PDCP}{Packet Data Convergence Protocol}
\newacronym{MAC}{MAC}{Medium Access Control}
\newacronym{RLC}{RLC}{Radio Link Control}
\newacronym{RRC}{RRC}{Radio Resource Control}
\newacronym{AM}{AM}{Acknowledged Mode}
\newacronym{UM}{UM}{Unacknowledged Mode}
\newacronym{TM}{TM}{Transparent Mode}
\newacronym{MIMO}{MIMO}{Multiple Input Multiple Output}
\newacronym{MISO}{MISO}{Multiple Input Single Output}
\newacronym{SIMO}{SIMO}{Single Input Multiple Output}
\newacronym{SISO}{SISO}{Single Input Single Output}
\newacronym{MCC}{MCC}{Mobile Country Code}
\newacronym{MNC}{MNC}{Mobile Network Code}
\newacronym{S-TMSI}{S-TMSI}{Shortened Temporary Mobile Subscriber Identity}
\newacronym{IMSI}{IMSI}{International Mobile Subscriber Identity}
\newacronym{DRB}{DRB}{Dedicated Radio Bearer}
\newacronym{GUMMEI}{GUMMEI}{Globally Unique MME Identity}

\newacronym{PCI}{PCI}{Physical-layer Cell Identity}
\newacronym{ROHC}{ROHC}{Robust Header Compression}
\newacronym{SN}{SN}{Sequence Number}
\newacronym{RAR}{RAR}{Random Access Response}
\newacronym{C-RNTI}{C-RNTI}{Cell Radio Network Temporary Identifier}
\newacronym{BSR}{BSR}{Buffer Status Report}
\newacronym{DRX}{DRX}{Discontinuous Reception}
\newacronym{PHR}{PHR}{Power Head Room}
\newacronym{PUSCH}{PUSCH}{Physical Uplink Shared Channel}
\newacronym{ADM}{ADM}{Activation/Deactivation MAC}
\newacronym{GP}{GP}{Gap Period}
\newacronym{CP}{CP}{Cyclic Prefix}
\newacronym{RE}{RE}{Resource Element}
\newacronym{RB}{RB}{Resource Block}
\newacronym{REG}{REG}{Resource Element Group}
\newacronym{CSRS}{CSRS}{Cell-Specific Reference Signal}
\newacronym{IFFT}{IFFT}{Inverse Fast Fourier Transform}
\newacronym{OFDMA}{OFDMA}{Orthogonal Frequency Division Multimple Access}
\newacronym{CRC}{CRC}{Cyclic Redundancy Check}

\newacronym{eNB}{eNB}{Evolved NodeB}
\newacronym{RAN}{RAN}{Radio Access Network}
\newacronym{ARQ}{ARQ}{Automatic Repeat reQuest}
\newacronym{NAS}{NAS}{Non-Access Stratum}
\newacronym{MME}{MME}{Mobility Management Entity}
\newacronym{MIB}{MIB}{Master Information Block}
\newacronym{SIB}{SIB}{System Information Block}
\newacronym{RSRP}{RSRP}{Reference Signal Received Power}
\newacronym{RAT}{RAT}{Radio Access Technologie}
\newacronym{ACK}{ACK}{Acknowledge}
\newacronym{NACK}{NACK}{Negative acknowledge}
\newacronym{PDCCH}{PDCCH}{Physical Downlink Control Channel}
\newacronym{SAW}{SAW}{Stop and Wait}
\newacronym{TTI}{TTI}{Transmission Time Interval}
\newacronym{RRH}{RRH}{Radio Remote Head}
\newacronym{SNIR}{SNIR}{Signal-to-Noise-plus-Interference Ratio}
\newacronym{WCET}{WCET}{Worst Case Execution Time}
\newacronym{GPC}{GPC}{General Purpose Computer}
\newacronym{KPI}{KPI}{Key Performance Indicator}
\newacronym{OAI}{OAI}{Open Air Interface}
\newacronym{IMS}{IMS}{IP Multimedia Subsystem}
\newacronym{vIMS}{vIMS}{virtual IP Multimedia Subsystem}
\newacronym{EPC}{EPC}{Evolved Packet Core}
\newacronym{SDN}{SDN}{Software Defined Network}
\newacronym{C-RAN}{C-RAN}{Centralized-RAN}
\newacronym{OS}{OS}{Operating System}
\newacronym{TB}{TB}{Transport Block}
\newacronym{TBS}{TBS}{Transport Block Size}
\newacronym{QCI}{QCI}{QoS Channel Indicator}
\newacronym{BER}{BER}{Bit Error Rate}
\newacronym{MEC}{MEC}{Multi-access Edge Computing}

\newacronym{GPU}{GPU}{Graphics Processing Unit}
\newacronym{CPU}{CPU}{Central Processing Unit}
\newacronym{ICIC}{ICIC}{Inter-Cell Interference Coordination}
\newacronym{SDU}{SDU}{Service Data Unit}
\newacronym{CBS}{CBS}{Code Block Size}
\newacronym{CB}{CB}{Code Block}
\newacronym{SPMD}{SPMD}{Single Program Multiple Data}
\newacronym{SIMD}{SIMD}{Single Instruction Multiple Data} 
\newacronym{IT}{IT}{Information Technology} 

\newacronym{SINR}{SINR}{Signal-to Interference Noise Ratio}
\newacronym{CO}{CO}{Central Office}
\newacronym{CA}{CA}{Carrier Aggregation}
\newacronym{SRS}{SRS}{Sound Reference Signal}
\newacronym{SC-OFDMA}{SC-OFDMA}{Single Carrier - Orthogonal Frequency Division Multiple Access}
\newacronym{FPGA}{FPGA}{Field-Programmable Gate Array}
\newacronym{TA}{TA}{Time Advancing}
\newacronym{CoMP}{CoMP}{Coordinated Multi-point}
\newacronym{NPRB}{NPRB}{Number of Physical Resource Blocks}
\newacronym{RTT}{RTT}{Round Trip Time}
\newacronym{CPRI}{CPRI}{Common Public Radio Interface}
\newacronym{CBR}{CBR}{Constant Bit Rate}
\newacronym{NRB}{NRB}{Number of Resource Blocks}
\newacronym{BJF}{BJF}{Biggest Job First}
\newacronym{EDF}{EDF}{Earliest Deadline First}
\newacronym{FCFS}{FCFS}{First-come, First-served}
\newacronym{PSTN}{PSTN}{Public Switched Telephone Network}
\newacronym{ETSI}{ETSI}{European Telecommunications Standards Institute}
\newacronym{vBBU}{vBBU}{virtualized BBU}
\newacronym{vRAN}{vRAN}{virtualized RAN}
\newacronym{IoT}{IoT}{Internet of Things}
\newacronym{B2B}{B2B}{Business to Business}
\newacronym{B2C}{B2C}{Business to Customer}
\newacronym{QoE}{QoE}{Quality of Experience}
\newacronym{QoS}{QoS}{Quality of Service}
\newacronym{VNO}{VNO}{Virtual mobile Network Operator}
\newacronym{SLA}{SLA}{Service Level Agreement}
\newacronym{VRRM}{VRRM}{Virtual Radio Resource Management}
\newacronym{KVM}{KVM}{Kernel-based Virtual Machine}
\newacronym{LXC}{LXC}{Linux Containers}
\newacronym{PS}{PS}{Processor Sharing}

\newacronym{eCPRI}{eCPRI}{evolved CPRI}
\newacronym{RoE}{RoE}{Radio over Ethernet}
\newacronym{PAPR}{PAPR}{Peak-to-average power ratio}
\newacronym{SC-FDMA}{SC-FDMA}{Single Carrier Frequency Division Multiple Access}
\newacronym{AGC}{AGC}{Automatic Gain Control}
\newacronym{PMD}{PMD}{Polarization Mode Dispersion}
\newacronym{ADC}{ADC}{Analogic-Digital Converter}

\newacronym{I/Q}{I/Q}{In-Phase Quadrature}

\newacronym{xRAN}{xRAN}{extensible Radio Access Network}
\newacronym{ISI}{ISI}{Inter-symbol interference}

\newacronym{FFT}{FFT}{Fast Fourier Transform}
\newacronym{IPC}{IPC}{Inter process communication}
\newacronym{CCDU}{CCDU}{Channel Coding Data Unit}
\newacronym{CC}{CC}{Channel Coding}
\newacronym{gNB}{gNB}{next-Generation Node B}
\newacronym{EUTRAN}{EUTRAN}{Evolved Universal Terrestrial Radio Access Network}
\newacronym{SCTP}{SCTP}{Stream Control Transmission Protocol}
\newacronym{NR}{NR}{New Radio}
\newacronym{NF}{NF}{Network Function}
\newacronym{CU}{CU}{Central Unit}
\newacronym{DU}{DU}{Distributed Unit}
\newacronym{NGC}{NGC}{Next Generation Core}
\newacronym{DL}{DL}{downlink}
\newacronym{UL}{UL}{uplink}
\newacronym{LJF}{LJF}{Largest Job First}
\newacronym{RANaaS}{RANaaS}{RAN as a Service}
\newacronym{NS}{NS}{Network Service}
\newacronym{FG}{FG}{Forwarding Graph}
\newacronym{VNFC}{VNFC}{VNF Component}

\newacronym{MANO}{MANO}{Management and Orchestration}
\newacronym{FIFO}{FIFO}{First In Firs Out}
\newacronym{NFVI}{NFVI}{NFV Infrastructure}
\newacronym{NFVO}{NFVO}{NFV Orchestrator}
\newacronym{PoP}{PoP}{Point of Presence}
\newacronym{NAT}{NAT}{Network Address Translation}
\newacronym{CDN}{CDN}{Content Delivery Network}
\newacronym{VNFM}{VNFM}{VNF Manager}
\newacronym{EM}{EM}{Element Management}
\newacronym{VIM}{VIM}{Virtualised Infrastructure Manager}
\newacronym{VMM}{VMM}{Virtual Machine Monitor}
\newacronym{e2e}{e2e}{end-to-end}
\newacronym{OTT}{OTT}{over-the-top}
\newacronym{ABI}{ABI}{Application Binary Interface}
\newacronym{API}{API}{Application Programing Interface}
\newacronym{ISA}{ISA}{Instruction Set Architecture}
\newacronym{JVM}{JVM}{Java Virtual Machine}
\newacronym{REST}{REST}{Representational State Transfer}
\newacronym{SOA}{SOA}{Service Oriented Architecture}

\newacronym{HSS}{HSS}{Home Subscriber Server}
\newacronym{Diah}{Diah}{Diameter handler}
\newacronym{ILP}{ILP}{Integer Linear Programming}
\newacronym{ISP}{ISP}{Internet Service Provider}
\newacronym{MIQCP}{MIQCP}{Mixed Integer Quadratically Constrained Program}
\newacronym{MILP}{MILP}{Mixed Integer Linear Programming}
\newacronym{CCO}{CCO}{Core Central Office}
\newacronym{MCO}{MCO}{Main Central Office}
\newacronym{ONAP}{ONAP}{Open Networking Automation Platform}
\newacronym{RAM}{RAM}{Random Access Memory}
\newacronym{PM}{PM}{Physical Machine}
\newacronym{DRF}{DRF}{Dominant Resource Fairness}
\newacronym{YARN}{YARN}{Yet Another Resource Negotiator}
\newacronym{DRFH}{DRFH}{DRF in Heterogeneous environments}
\newacronym{FQ}{FQ}{Fair Queuing}
\newacronym{GPS}{GPS}{Generalized  Processor Sharing}
\newacronym{WFQ}{WFQ}{Weighted Fair Queuing}